\documentclass[12pt]{article}

\usepackage{a4wide}
\usepackage{amsmath}
\usepackage{amsfonts}
\usepackage{amsthm}
\usepackage{dsfont}

\newcommand{\Int}{\int\limits}

\newcommand{\ud}{\mathrm{d}}
\newcommand{\ui}{\mathrm{i}}
\newcommand{\ue}{\mathrm{e}}

\newcommand{\R}{\mathds{R}}
\newcommand{\Z}{\mathds{Z}}
\newcommand{\N}{\mathds{N}}

\newcommand{\cU}{{\mathcal U}}
\newcommand{\cA}{{\mathcal A}}
\newcommand{\cH}{{\cal H}}

\newcommand{\supp}{\operatorname{supp}}

\providecommand{\norm}[1]{\lVert#1\rVert}
\providecommand{\abs}[1]{\lvert#1\rvert}
\providecommand{\biggabs}[1]{\biggl\lvert#1\biggr\rvert}

\newcommand{\pa}{\partial}
\newcommand{\la}{\langle}
\newcommand{\ra}{\rangle}

\newcommand{\bT}{\mathbf{T}}
\newcommand{\Op}{\operatorname{Op}}

\newcommand{\rank}{\operatorname{rank}}

\newtheorem{theorem}{Theorem}
\newtheorem{lemma}{Lemma}
\newtheorem{proposition}{Proposition}

\newtheorem*{Condition}{Condition}{\bf}{\rm}

\begin{document}
\title{Semiclassical behaviour of expectation values in 
time evolved Lagrangian states 
for large times}

\author{Roman Schubert\thanks{Department of Mathematics, University of Bristol, UK, e-mail: roman.schubert@bristol.ac.uk}}


\date{January 30, 2004}

\maketitle

\begin{abstract}
We study the behaviour of time evolved quantum mechanical expectation values in
Lagrangian states in the limit $\hbar\to 0$ and $t\to\infty$. 
We show that it 
depends strongly on the dynamical properties of 
the corresponding classical system. If the classical system is 
strongly chaotic, i.e. Anosov, then the expectation values tend to a 
universal limit. This can be viewed as an analogue of mixing in the 
classical system. If the classical system is integrable, then 
the expectation values need not converge, and if they converge 
their limit depends on the initial state. An additional 
difference occurs in the timescales for which we can prove this behaviour, 
in the chaotic case we get up to Ehrenfest time, 
$t\sim \ln(1/\hbar)$, whereas for integrable system we have a much larger 
time range.   
\end{abstract}

\section{Introduction and results}

A striking property of chaotic dynamical systems is the universality 
which these systems show in the time evolution for large times. 
Let $(\Sigma,\Phi^t, \ud \mu)$ be a dynamical system, i.e., 
$\Sigma$ is the compact phase space, $\Phi^t:\Sigma\to\Sigma$ the flow and 
$\ud \mu$ a  normalised invariant measure on $\Sigma$. If the system is 
mixing
then for any 
$\rho,a\in L^2(\Sigma,\mu)$ with $\int \rho\,\, \ud \mu =1$ one has 
\begin{equation}\label{eq:mixing}
\int a\circ\Phi^t \, \rho \,\, \ud \mu \to \int a\,\, \ud\mu \,\, ,\quad 
\text{for}\,\,  t\to\infty\,\, .
\end{equation}
If we think of $\rho$ as describing an probability distribution of initial 
states and of $a$ as an observable, then mixing means that the system 
forgets its initial conditions for large times and so one needs only 
to know the ``equilibrium state'' $\ud \mu$ in order to predict 
the behaviour of time evolved observables 
for large times. If the rate of mixing is fast enough this then often implies 
other universal statistical features, e.g., a central limit theorem 
for time means of observables.

We want to explore to what extend this universality shows up in 
quantum mechanics, too. The analogue of the expectation value in 
\eqref{eq:mixing} is a quantum mechanical expectation value for a time 
evolved state. So let $\cU(t)$ denote the time evolution operator 
of our quantum system,  
$\cA$ an observable, i.e. a bounded operator, and $\psi$ a state, 
we want to know if 
\begin{equation}
\la \cU(t)\psi, \cA\cU(t)\psi\ra 
\end{equation}
converges to some limit if $\hbar\to 0$ and $t\to\infty$, at least 
for certain classes of observables and states. 
We will consider here Lagrangian states as initial states 
and bounded pseudo-differential operators as observables.

The main difficulty in this problem comes from the fact that we have to 
perform two limits, $\hbar\to 0$ and $t\to\infty$, and these two limits 
do not commute. So we have to specify precisely how we take the joint limit 
and we have to use semiclassical constructions which are 
to some extend uniform in $t$. For systems which have some positive 
Liapunov exponents, it was found in the late 70's in the physics literature 
 \cite{BerZas78,Zas81,BerBalTabVor79,BerBal79},  that the usual 
semiclassical constructions apparently can only  work up to a timescale 
which grows  
logarithmically in $\hbar$, $T_E\sim \ln(1/\hbar)$, the so called Ehrenfest 
or log-breaking time. That semiclassical constructions actually 
do work up to that time was rigorously proved in \cite{ComRob97} for the 
time evolution of coherent states and in \cite{BamGraPau99} for the 
time evolution of observables. We will use for our work the results 
in \cite{BouRob02} who extended  the results by Bambusi, Graffi 
and Paul.

The time range beyond the Ehrenfest time is not well understood yet. 
But results by 
Tomsovic,  
Heller and coworkers, \cite{TomHel91,TomHel93,OCoTomHel92},  
suggest that semiclassical methods might be extended beyond Ehrenfest time. 
They studied  for autocorrelation functions of coherent states 
the question if one can extend the semiclassical propagator to 
timescales which are 
algebraic in $1/\hbar$, and  demonstrated  numerically that this is 
possible for the stadium billiard and some quantised maps.  

One motivation for this  work are the results of Bonechi and de Bievre 
for the time evolution of coherent states in cat-maps, \cite{BondeB00}. 
They showed that in the cat-map a time evolved coherent state 
becomes equidistributed just after the Ehrenfest time, but 
they could control the time evolution only up to a slightly larger time 
range which is still logarithmic in $1/\hbar$.  
But since one expect a coherent state to become stretched along 
the unstable manifold of the orbit on which it is centred, 
it might be effectively modelled by a Lagrangian state associated with 
this unstable manifold. This is one motivation for studying Lagrangian 
states. More recently estimates on the time evolution around Ehrenfest 
time have been used in \cite{deBFauNon03} to construct scared eigenstates 
for the quantised cat map, and in \cite{deBRob03} the time evolution of 
coherent states along the seperatrix in one-dimensional 
systems was investigated.

A typical Lagrangian state on a manifold $M$ is of the form 
\begin{equation}\label{eq:def-of-lagrangian-state}
\psi(x)=\rho(\hbar,x)\ue^{\frac{\ui}{\hbar} \varphi(x)}\,\, ,
\end{equation}
where $\varphi$ is a smooth real valued function and 
$\rho(\hbar,x)$ is a smooth function with compact support 
with an asymptotic expansion 
$\rho(\hbar,x)\sim \rho_0(x)+\hbar\rho_1(x)+\cdots $ 
for $\hbar\to 0$. 
The important geometrical object associated with $\psi$ is 
the Lagrangian manifold generated by the phase function $\varphi$, 
\begin{equation}\label{eq:lagrange-gen-phi}
\Lambda_{\varphi}:= \{ (\varphi'(x),x)\,\, ;x\in U\}\,\, .
\end{equation}
We will denote the set of these states with compact support by $I_0(\Lambda)$. The definition 
can be extended to arbitrary Lagrangian manifolds, i.e., they 
need not be representable in the form 
\eqref{eq:lagrange-gen-phi}. Any Lagrangian submanifold $\Lambda\subset T^*M$ 
can be represented locally as 
$\Lambda=\{ (\varphi'_x(x,\theta),x)\,; \, \varphi'_{\theta}(x,\theta)=0, \, 
(x,\theta)\in U\times \R^{\kappa}\}$, where 
$\varphi(x,\theta)$ is non-degenerate, i.e., 
$\rank(\varphi''_{x,x}(x,\theta),\varphi''(x,\theta)(x,\theta))=d$ for 
$(x,\theta)$ with $\varphi'_\theta(x,\theta)=0$. The corresponding 
Lagrangian states are given by  
\begin{equation}
\psi(x)=\frac{1}{(2\pi\hbar)^{\kappa/2}}\int_{\R^{\kappa}} 
\rho(\hbar, x,\theta )\ue^{\frac{\ui}{\hbar} \varphi(x,\theta)}\,\, 
\ud \theta\,\, ,
\end{equation}
see \cite{Dui74,BatWei97} and 
\cite[Section 1.2.1]{Ivr98} for more details. 
Lagrangian states appear quite often in applications, 
e.g., if $\varphi(x)=\la p,x\ra$ we have a localised plane wave with 
momentum $p$ 
or if $\varphi$ depends only on $\abs{x}$ we get circular waves. 
Since the simultaneous eigenstates of $d$ commuting 
pseudo-differential operators are typically Lagrangian, 
this class of states appears quite frequently as the result of 
the preparation of an experiment, e.g., the above mentioned 
examples occur if one selects initial states with 
certain momentum, or certain angular momentum, respectively.

The leading order behaviour of a Lagrangian state 
$\psi$ for $\hbar\to 0$ is determined by its 
principal symbol $\sigma (\psi)$ which, modulo phase factors, is 
a half-density on $\Lambda$. In the case that 
$\psi$ is of the form \eqref{eq:def-of-lagrangian-state} 
$\sigma(\psi)$ is the pullback of the half-density 
$\rho_0(x) \abs{\ud x}^{1/2}$ 
on $\R^d$ by the projection $\pi :\Lambda_{\varphi}\to \R^d$. 
We will only encounter its modulus squared, 
the density $\abs{\sigma (\psi)}^2$, which can be defined more directly  by 
the relation
\begin{equation}
\int_{\Lambda} a\, \abs{\sigma(\psi)}^2:= 
\int_{\R^d} a(\varphi'(x),x) \abs{\rho_0(x)}^2 \,\, \ud x
\end{equation}
for any $a\in C^{\infty}(T^*M)$.

The observables we consider are given by pseudo-differential operators. 
We will say that $\cA\in\Psi^m(M)$ if 
locally $\cA=\Op[a]$ where 
\begin{equation}\label{eq:weyl-quant}
\Op[a]\psi(x)=\frac{1}{(2\pi\hbar)^d}\iint_{\R^d\times \R^d} a\big(\frac{x+y}{2},\xi\big) \ue^{\frac{\ui}{\hbar} \la x-y,\xi\ra} \psi(y)\,\, \ud y\, \ud \xi\,\, , 
\end{equation}
and the symbol $a(\hbar,x,\xi)$ has an asymptotic expansion 
$a(\hbar,x,\xi)\sim a_0(x,\xi)+\hbar a_1(x,\xi)+ \hbar^2 a_2(x,\xi)+\cdots$ and satisfies 
\begin{equation}
\abs{\pa_x^{\alpha}\pa_{\xi}^{\beta}a(\hbar,x,\xi)}\leq 
C_{\alpha,\beta} (1+\abs{x}^2+\abs{\xi}^2)^{m/2}\,\, , 
\end{equation}
for $\hbar\in (0,1]$ and all $\alpha,\beta\in\N^d$. 
One calls 
$\sigma(a):=a_0$ the principal symbol of $a$, or of $\cA$,, and although 
the full symbol $a$ is only defined locally, the principal symbol 
defines a function on $T^*M$, i.e., on phase space.  
The operators in 
$\Psi^{0}(M)$ are bounded, and they will form the set of observables for 
which we study time evolution. See e.g. \cite{DimSjo99}  for more details.

Our first assumption on the system is 
that the Hamiltonian fits into the above framework, i.e., is 
a pseudo-differential operator. 

\begin{Condition}[H]
 Let $M$ be a $C^{\infty}$ manifold and let $\cH \in \Psi^m(M)$, 
for some $m\in\R$,  be essentially selfadjoint.  
\end{Condition}

A typical example is $\cH=-\hbar^2\Delta_g+V$, where $\Delta_g$ is the 
Laplace Beltrami operator associated with a metric $g$ on $M$, and 
$V$ is a smooth real valued function (with $\abs{\pa^{\alpha}V(x)}\leq C_{\alpha}(1+\abs{x})^m$ if $M$ is not compact). For conditions on general operators from 
$\Psi^m(M)$ to be (essentially) selfadjoint see \cite{DimSjo99}.  

The Hamiltonian flow on $T^*M$ generated by the principal symbol 
$H$ of 
$\cH$ will be denoted by $\Phi^t$. 

\begin{Condition}[O]
There exists an open connected set 
$\Omega \subset T^*M$ which has compact closure  
and which is invariant under the flow $\Phi^t$.
\end{Condition}

Let $\Sigma_E:=\{ z\in T^*M\, ;\, H_0(z)=E\}$ be the energy shell of 
energy $E$ 
and denote by $\ud \mu_E$ the Liouville measure on $\Sigma_E$. 
$\Sigma_E$ and $\ud\mu_E$ are  invariant under the flow. Let us recall the 
definition of an Anosov flow, 

\begin{Condition}[A]
A flow $\Phi^t$ on a compact manifold $\Sigma$ is  called Anosov, if 
for every $x\in \Sigma$ there exists a splitting 
$T_x\Sigma=E^s(x)\oplus E^u(x)\oplus E^0(x)$ which is invariant 
under $\Phi^t$ and where $E^0(x)$ is one-dimensional and 
spanned by the generating vectorfield of $\Phi^t$. 
Furthermore there exist constants $C,\lambda >0$ such that 
\begin{align}
||\ud \Phi^t v ||& \leq C \ue^{-\lambda t} ||v|| \qquad 
\text{for each  $ v\in E^s $ and $t\geq 0$} \\
||\ud \Phi^t v ||& \leq C \ue^{\lambda t} ||v|| \qquad 
\text{for each  $ v\in E^u $ and $t\leq 0$}\,\, .
\end{align}
\end{Condition}

The two distributions  $E^s$ and $E^u$ can be integrated to give the 
stable and unstable foliations, respectively. We will denote the leaves
through $x$ by $W^s(x)$ and $W^u(x)$. If the flow is smooth then 
the leaves are smooth submanifolds but the dependence of the leaves on 
$x$ is usually only H{\"o}lder continuous, and we will denote the 
H{\"o}lder exponent by $\alpha$. The corresponding 
weakly stable and unstable manifolds are defined by 
$W^{ws/wu}(x):=\bigcup_{t\in\R} \Phi^t(W^{s/u}(x))$. 
If $\Sigma$ is an energy-shell of an Hamiltonian system, and 
$\Phi^t$ the Hamiltonian flow, then $W^s(x)$ and $W^u(x)$ have the 
same dimension, and $W^{ws}(x)$ and $W^{wu}(x)$ are Lagrangian submanifolds.

An example for an Anosov flow is given by the geodesic flow on a compact 
manifold of negative curvature, see e.g. \cite{Ebe01}. If the 
Hamilton operator is the Laplace Beltrami operator associated with such a 
metric, then the flow generated by the principal symbol of this operator is 
conjugate to  the geodesic flow, and its restriction to any 
equi-energy shell $\Sigma_E$ is Anosov.

For the time evolution of the Lagrangian states the position of $\Lambda$ 
relative to the stable foliation will be important. Namely we have to require 
that $T_x\Lambda$ contains no stable directions for most 
$x$, this  leads to the following transversality conditions. 

\begin{Condition}[T]
\begin{itemize}
\item[(i)] If $\Lambda\in \Sigma_E$ then assume that 
$T_x\Lambda\cap E^s(x)=\{0\}$ for all 
$x\in \Lambda\backslash \Lambda_{sing}$, where 
$\Lambda_{sing}\subset \Lambda$ has at least codimension $1$. 
\item[(i)] If $\Lambda\subset\Omega$ and the flow is Anosov 
on all $\Sigma_E\subset\Omega$ then for all such $E$ assume that  
$T_x(\Lambda\cap\Sigma_E)\cap (E^s(x)\oplus E^0(x))=\{0\}$ for all 
$x\in (\Lambda\cap\Sigma_E)\backslash \Gamma_{E,sing}$, where 
$\Gamma_{E,sing}\subset (\Lambda\cap\Sigma_E)$ has at least codimension 
$1$.
\end{itemize}
\end{Condition}

These conditions are generically fulfilled, so a typical 
Lagrangian manifold $\Lambda$ will satisfy them.  
We can state now the main result of this paper about 
expectation values of time evolved Lagrangian states.


\begin{theorem}\label{thm:main-anosov}
Let $M$ be a $C^{\infty}$ manifold, 
and $\cH\in \Psi^m(M)$ be a selfadjoint pseudo-differential 
operator on $M$, with principal symbol $H_0$. Let $\Phi^t$ be the Hamiltonian 
flow on $T^*M$ generated by $H_0$, and assume condition 
(O) is fulfilled. Let $\Lambda \subset \Omega$ 
be a Lagrangian submanifold. Then 

\begin{itemize}
\item[(i)] if $\Lambda\subset \Sigma_E\subset \Omega$,  the flow on 
$\Sigma_E$ is Anosov, and $\Lambda$ satisfies condition (T)(i), then there exist 
for every $\psi\in I_0(\Lambda)$ and 
$\Op[a]\in \Psi^0(M)$ constants $C,c,\Gamma, \gamma>0$ such that 
\begin{equation}\label{eq:lambda1-limit}
\bigg|\la \cU(t) \psi , \Op[a]\cU(t) \psi \ra -\int_{\Sigma_E}\sigma(a)\,\, \ud\mu_E
\int_{\Lambda}\, \abs{\sigma(\psi)}^2\bigg|
\leq C\hbar \ue^{\Gamma \abs{t}}+c\ue^{-\gamma t}
\end{equation}
\item[(ii)] if the flow is Anosov on all $\Sigma_E\subset \Omega$, and 
$\Lambda\cap \Sigma_E$ satisfies condition (T)(ii), then 
there exist for every $\psi \in I_0(\Lambda)$ and 
$\Op[a]\in \Psi^0(M)$ constants $C,c,\Gamma,\gamma$ such that 
\begin{equation}\label{eq:lambda2-limit}
\bigg|\la \cU(t) \psi , \Op[a]\cU(t) \psi \ra -\int\Int_{\Sigma_E}\sigma(a)\,\, \ud\mu_{E}
\Int_{\Lambda\cap\Sigma_E}\, \abs{\sigma(\psi )}_E^2\, \, \ud E\bigg|
\leq C\hbar \ue^{\Gamma \abs{t}}+c\ue^{-\gamma t}\,\, ,
\end{equation}
where the density $\abs{\sigma(\psi)}_E^2$ on $\Lambda\cap \Sigma_E$ is defined by $\abs{\sigma(\psi)}^2=\abs{\sigma(\psi)}_E^2\otimes\abs{\ud E}$. 
\end{itemize}
\end{theorem}


In order that the right hand sides of the inequalities \eqref{eq:lambda1-limit} and 
\eqref{eq:lambda2-limit} tend to zero for $\hbar \to 0$ and $t\to\infty$, we have to have 
\begin{equation}
t<< \ln(1/\hbar)\,\, ,
\end{equation}
so up to Ehrenfest time we get convergence. 

Let us compare this result with mixing for the classical system. To this 
end assume that $\norm{\psi}=1$, this implies that 
$\int_{\Lambda}\abs{\sigma(\psi)}^2 =1$ and then \eqref{eq:lambda1-limit} 
gives 
\begin{equation}
\la \cU(t)\psi, \Op[a]\cU(t)\psi\ra \to \int_{\Sigma_E}\sigma(a)\ud\mu_E
\end{equation}
for $t\to\infty$ and $\hbar\to 0$ such that $\hbar\ue^{\Gamma \abs{t}}\to 0$. 
So we have the same behaviour as in the classical system, see 
\eqref{eq:mixing}, in particular we obtain the same kind of universality. 
The limit does not depend any longer on the initial state as long as it 
satisfies the conditions of part (1) of Theorem \ref{thm:main-anosov}.

The transversality condition on the Lagrangian manifold is necessary. 
If $\Lambda$ 
is for instance the stable manifold of an periodic orbit $\gamma$, 
then one has for $\psi\in I_0(\Lambda)$
\begin{equation}\label{eq:per-orbit}
\la \cU(t)\psi ,\Op[a]\cU(t)\psi\ra 
=\sum_{k\in \Z} b_k \ue^{\frac{2\pi \ui }{T_{\gamma}} k t}
+O(\hbar \ue^{\Gamma\abs{t}}) +O(\ue^{-\gamma t}) 
\end{equation}
where $T_{\gamma}$ is the period of he orbit, and the coefficients $b_k$ are
related to $\sigma(\psi)$ and $\sigma(a)$. We will discuss 
this in more detail in Section \ref{sec:mixing}. 

The result in Theorem \ref{thm:main-anosov} can be viewed as an
analogue for time evolution of the quantum ergodicity results for 
eigenfunctions 
\cite{Shn74,Zel87,Col85,HelMarRob87}. 
If the classical system is ergodic then almost all eigenfunctions 
become equi-distributed. Here we obtain equidistribution under 
time evolution, but we need stronger conditions on the classical system. 
The main open problem now is to try to extend the time range in 
Theorem \ref{thm:main-anosov}. This could then in turn be used to 
improve the quantum ergodicity results for eigenfunctions.

We want to compare now the behaviour found in classically chaotic systems with 
integrable systems. Following \cite{BouRob02} we introduce the following 
integrability condition.

\begin{Condition}[I]
$M$ is analytic, and there exists a symplectic map $\chi$ from $\Omega$ 
into $U\times 
\mathbf{T}^d$ , where $U$ is an open set in $\R^d$ and $\mathbf{T}^d$ is an
d-dimensional torus such that  
\begin{equation}
\chi(\Phi^t(z))=(I(z), \varphi(z)+t\omega(I(z)))\, , \qquad \forall z\in\Omega, \forall t\in\R\, ,
\end{equation}
where $\chi(z)=(I(z),\varphi(z))$. Moreover there exists complex open neighbourhoods 
$\tilde{\Omega}$, $\tilde{U}$, $\tilde{\mathbf{T}}^d$ of $\Omega$, $U$, 
$\mathbf{T}^d$ such that $\chi$ is an analytic diffeomorphism from 
$\tilde{\Omega}$ onto $\tilde{U}\times \tilde{\mathbf{T}}^d$.  
\end{Condition} 

According to the Liouville Arnold Theorem this situation occurs 
if one has $d$ analytic integrals of motion which are in involution 
and which are independent on $\Omega$.

In the case of integrable systems one can explore larger time scales, and we obtain 
the following result.

\begin{theorem}\label{thm:limit-int-syst}
Assume conditions (H), (O) and (I) are  fulfilled and $\Lambda\subset \Omega$. Then 
there exists $C>0$ and $\beta>0$ such that  
 \begin{itemize}
\item[(i)] if $\Lambda$  is an invariant torus with frequency $\omega\in\R^d$  
we have for all $\psi\in I_0(\Lambda)$ 
\begin{equation}\label{eq:int-lambda-torus} 
\biggabs{\la \cU(t)\psi , \Op[a]\cU(t)\psi \ra -
\sum_{m\in\Z^d} \sigma(a)_m(I)\big(\abs{\sigma(\psi)}^2\big)_{-m} \ue^{\ui \la m,\omega(I)\ra\, t}}
\leq C \hbar(1+\abs{t})^{\beta} \,\, ,
\end{equation}
where $\sigma(a)_m(I)$ and $\big(\abs{\sigma(\psi)}^2\big)_{-m}$ are the Fourier-coefficients 
of $\sigma(a)(I,\cdot)$ and $\abs{\sigma(\psi)}^2$, respectively.
\item[(ii)] if the the system is non-degenerate, i.e., $\omega'(I)\neq 0$ on 
$U$, and $\Lambda$ is transversal to the foliation into invariant tori, then we have for all $\psi\in I_0(\Lambda)$
\begin{equation}\label{eq:int-lambda-transversal} 
\biggabs{\la \cU(t)\psi , \Op[a]\cU(t)\psi \ra-\int \sigma(a) \,\, \mu_{\psi, T} }
\leq  C \hbar(1+\abs{t})^{\beta}+c\frac{1}{1+\abs{t}}\,\, , 
\end{equation}
with $\mu_{\psi,\mathcal{T}}=\abs{\sigma(\psi)}^2\otimes \abs{\ud x}$. 
\end{itemize}
\end{theorem}

The density $\mu_{\psi,\mathcal{T}}=\abs{\sigma(\psi)}^2\otimes \abs{\ud x}$ 
can be described more explicitly in action angle coordinates. By the 
transversality assumption there exist local symplectic coordinates 
$(I,x)\subset U\times V$ such that $\Lambda=\{(I,0)\, ,\, I\in U\}$ 
and the sets $\{(I_0,x)\, ,\, x\in V\}$ belong to invariant tori. 
In this coordinates the modulus square of the principal symbol 
can be written as $\abs{\sigma(\psi)}^2=\abs{\hat{\rho}(I)}^2\abs{\ud I}$ 
and we get 
$\mu_{\psi,\mathcal{T}}=\abs{\hat{\rho}(I)}^2\abs{\ud I\wedge \ud x}$. 
So integrating against this density means that we take the mean over each 
invariant torus, and then integrate these contributions 
weighted with the principal symbol of the state.  This means that 
the knowledge of the limit density $\mu_{\psi,\mathcal{T}}$ 
allows to determine the foliation into invariant tori, and the 
distribution of the mass 
of the initial state across the tori.

In case of a chaotic system the situation is different. 
The only information on the initial state 
which survives is 
the information on how its mass is distributed among the energy shells. 
All other information 
is lost, and so we have the same degree of universality as in the 
classical system.

The organisation of the paper is as follows. In Section 
\ref{sec:redu_to_class} we reduce the quantum mechanical problem to 
one in classical mechanics, here the limitations on the time range occur. 
In Section \ref{sec:mixing} we extend previous results on mixing 
in Anosov systems and use them to prove Theorem \ref{thm:main-anosov}. 
In Section 
\ref{sec:integrable} we discuss the integrable case and give 
the proof of Theorem \ref{thm:limit-int-syst}.

\section{Reduction to classical dynamics}\label{sec:redu_to_class}

Our aim in this section is to reduce the quantum mechanical problem to 
a problem in classical mechanics. This is obtained in 

\begin{proposition}\label{prop:reduction} 
Assume the conditions (H) and (O), and let $\Lambda\subset \Omega$ be a 
Lagrangian manifold, $\psi \in I_0(\Lambda)$ and 
$\Op[a]\in\Psi^0(M)$. 
Then there exists a constant $\Gamma >0$, independent of $\Lambda$ and 
$a$, and $C>0$ such that
\begin{equation}\label{eq:red-gen}
\biggabs{\la \cU(t) \psi, \Op[a] \cU(t) \psi\ra  -\int_{\Lambda} \sigma(a)\circ\Phi^t \, 
\abs{\sigma(\psi)}^2} \leq C\hbar \ue^{\Gamma \abs{t}}\,\, .
\end{equation}
When condition (I) is fulfilled in addition then there exists a constant 
$\beta>0$ and $C>0$ 
such that
\begin{equation}\label{eq:red-int}
\biggabs{\la \cU(t) \psi, \Op[a] \cU(t) \psi\ra  -\int_{\Lambda} \sigma(a)\circ\Phi^t \, 
\abs{\sigma(\psi)}^2} \leq C\hbar (1+\abs{t})^{\beta} \,\, .
\end{equation}
\end{proposition}

The first step in the proof of this proposition is the following simple lemma. 
 
\begin{lemma}\label{lem:exp_value}
Let $\psi\in I_0(\Lambda)$ be a Lagrangian state with compact support on $M$, 
then there exists $C>0$ and an integer $k>0$ such that for all 
$\Op[a]\in\Psi^0(M)$
\begin{equation}
\bigg|\la \psi,\Op[a] \psi \ra -\int_{\Lambda} \sigma(a)\, 
|\sigma(\psi)|^2 \bigg|\leq C 
\sum_{\abs{\beta}\leq k} \abs{\pa^{\beta}a}\, \hbar
\end{equation}
\end{lemma} 

This is a standard result which follows from the results about application 
of pseudo-differential operators on Lagrangian states,  
see e.g. \cite{Hoe85b,BatWei97}, 
we have only made the dependence on $a$ 
of the right hand side more explicit. Since this Lemma 
is an application of the method of stationary phase, the remainder 
follows from the remainder estimates in this method, see \cite{Hor90}.

The second ingredient in the proof of Proposition \ref{prop:reduction}  
is an Egorov theorem which is valid up to  Ehrenfest time. 
The problem of time evolution of observables 
with remainder estimates uniform in time has been studied by 
Ivrii and Kachalkina in \cite[Chapter 2.3]{Ivr98}. Independently 
\cite{BamGraPau99} obtained  a proof 
of the validity of Egorov up to Ehrenfest time for analytic 
observables and Hamiltonians.  These results were then 
extended  in the 
work of Bouzouina and Robert, \cite{BouRob02}. 
In the formulation of the result we need the notion of essential support of 
an operator $\Op[a]\in \Psi^0(M)$. Recall that $z\in T^*M$ is not 
in the essential support of $\Op[a]$ if there is a neighbourhood 
$U$ of $z$ such that $\abs{a(z)}\leq C_N \hbar^N$ for all $N\in\N$ and 
$z\in U$. So $\Op[a]$ is semiclassically negligible outside of 
its essential support.


\begin{theorem}[\cite{BouRob02}]\label{thm:egorov}
Assume the conditions (H) and (O). 
Then there  exists a constant $\Gamma_1 >0$ such that 
for any $\Op[a]\in \Psi^0(M)$ with essential support in $\Omega$
there is a $C>0$ such that 
\begin{equation}
\norm{\cU(t)^*\Op[a]\cU(t)-\Op[a\circ\Phi^t]}\leq C \hbar \ue^{\Gamma_1 t}\,\, .
\end{equation}
\end{theorem}  


A much stronger version of this theorem was proved for 
$M=\R^n$ in \cite{BouRob02}, but the generalisation of their result 
to manifolds is complicated since the higher order terms of 
the symbol are not invariantly defined on $T^*M$. 
But we only need the leading order term, i.e. the principal symbol, 
and since this is a function on $T^*M$ the result generalises to the case 
of manifolds.

In case of integrable systems we will use instead the stronger 
Theorem 1.13 from 
\cite{BouRob02}. 


\begin{theorem}[\cite{BouRob02}]\label{thm:egorov-int}
Assume conditions (H), (O) and (I), then for every $\Op[a]\in\Psi^0(M)$ 
with essential support in $\Omega$ 
there exist constants $C>0$ and $\beta_d\leq 5d+4$ such that 
\begin{equation}
\norm{\cU(t)^*\Op[a]\cU(t)-\Op[a\circ\Phi^t]}\leq C\hbar (1+\abs{t})^{\beta_d}\,\, .
\end{equation}
\end{theorem} 


We can now conclude the proof of Proposition \ref{prop:reduction}. 

\begin{proof}[Proof of Proposition \ref{prop:reduction}] 
We will first assume that the essential support 
of $\Op[a]$ is contained on $\Omega$. 
Then by Theorem \ref{thm:egorov} we have that 
\begin{equation}
\abs{\la\cU(t)\psi,\Op[a]\cU(t)\psi\ra-\la\psi,\Op[a\circ\Phi^t]\psi \ra}
\leq C\hbar \ue^{\Gamma_1 \abs{t}}\,\, ,
\end{equation}
and Lemma \ref{lem:exp_value} gives 
\begin{equation}
\biggabs{\la\psi,\Op[a\circ\Phi^t]\psi,\ra-\int_{\Lambda} \sigma(a)\circ\Phi^t \abs{\sigma(\psi)}^2} \leq C \sum_{\abs{\beta}\leq k}\abs{\pa^{\beta} (a\circ\Phi^t)}\,\, \hbar\,\, .
\end{equation}
But as is well known, 
$\sum_{\abs{\alpha}\leq k}\abs{\pa^{\alpha}(a\circ\Phi^t)}\leq C \ue^{\Gamma_2\abs{t}} 
\sum_{\abs{\alpha}\leq k}\abs{\pa^{\alpha}a}$ for some $\Gamma_2>0$, see e.g. 
\cite[Lemma 2.4]{BouRob02}, and combining these estimates gives 
\eqref{eq:red-gen} with $\Gamma=\max\{\Gamma_1,\Gamma_2\}$. 
For the proof of equation \eqref{eq:red-int} we use Theorem 
\ref{thm:egorov-int} together with Lemma \ref{lem:exp_value} 
to get 
\begin{equation}
\biggabs{\la\cU(t)\psi,\Op[a]\cU(t)\psi\ra-\int_{\Lambda} \sigma(a)\circ\Phi^t \abs{\sigma(\psi)}^2}\leq C\hbar(1+\abs{t})^{\beta_d} +C' \sum_{\abs{\alpha}\leq k}\abs{\pa^{\alpha}(a\circ\Phi^t)}\,\,  \hbar 
\end{equation}
and with the estimate 
$\sum_{\abs{\alpha}\leq k}\abs{\pa^{\alpha}(a\circ\Phi^t)}\leq C'' 
\sum_{\abs{\alpha}\leq k}\abs{\pa^{\alpha}a} (1+\abs{t})^{\beta_d'}$, see 
\cite[Lemma 4.2]{BouRob02},  
the proof is complete 
if we take $\beta=\max\{\beta_d,\beta_d'\}$.

We finally show that we can reduce the case of an arbitrary 
observable $\Op[a]\in \Psi^0(M)$ to the case  of 
observables with essential support in $\Omega$. 
Let $I_0:=H_0^{-1}(\Omega)$ and $I_1:=H_0^{-1}(\supp(\sigma(\psi)))$, 
where $H_0$ is the principal symbol of $\cH$, 
be the energy-ranges of $\Omega$ and the support 
of $\sigma(\psi)$ on $\Lambda$, respectively. 
Then $I_0$ is an open interval, $I_1$ is a closed interval with 
$I_1\subset I_0$, and so there exists a function 
$f\in C_0^{\infty}(I_0)$ with $f|_{I_1}\equiv 1$. 
Then by the functional calculus, see \cite{DimSjo99}, 
the operator $f(\cH)$ is in $\Psi^0(M)$, has  essential support in 
$\Omega$,  commutes with $\cU(t)$, and satisfies 
$\norm{f(\cH)\psi-\psi}\leq C\hbar$. 
Therefore 
\begin{equation}
\abs{\la \cU(t)\psi, \Op[a]\cU(t)\psi\ra- \la \cU(t)\psi, f(\cH)\Op[a]\cU(t)\psi\ra}\leq C\hbar\, \, ,
\end{equation}
and since the essential support of $f(\cH)\Op[a]$ is contained in 
$\Omega$ we are done. 
\end{proof}

\section{Chaotic systems}\label{sec:mixing}

By Proposition \ref{prop:reduction} the proof of Theorem \ref{thm:main-anosov}
is now reduced to the study of 
\begin{equation}
\int_{\Lambda}\sigma(a)\circ\Phi^t\, \abs{\sigma(\psi)}^2
\end{equation}
and this expression is very similar to a correlation function like 
in \eqref{eq:mixing}. The only difference is that the density 
$\rho$ is replaced by a density concentrated on the submanifold $\Lambda$. 
Our aim in this section is to extend existing results on mixing 
of Anosov flows to this modified correlation functions. 
It is clear that we need a condition on the manifold $\Lambda$, 
as the example of a weakly stable manifold shows. Because if 
$\Lambda$ is the weakly stable manifold of a periodic 
trajectory, then the mass of 
$a$ will become more and more concentrated on that trajectory and will 
not become equidistributed. This example will be discussed in more 
detail  at the end of this section.

Recall that a function $a$ on a set $X$ with metric $d(x,y)$ 
is H{\"o}lder continuous with H{\"o}lder exponent $\alpha\in(0,1)$ if 
$\abs{a(x)-a(y)}\leq C d(x,y)^{\alpha}$ and the smallest 
constant $C$ is called it 
H{\"o}lder constant $\abs{a}_{\alpha}$. The set of H{\"o}lder continuous functions on a set $X$ will be denoted by $C^{\alpha}(X)$. 
Following the usual conventions we will fix a metric on  
the energy shell $\Sigma_E$, which then in turn induces metrics on  
submanifolds of $\Sigma_E$.

We will rely mainly on Liverani's recent result on mixing 
for contact Anosov flows, \cite{Liv03}. He shows that 
for any $\alpha\in (0,1)$ there exist constants $C,\gamma>0$ 
such that for $a,b\in C^{\alpha}(\Sigma)$ one has
\begin{equation}\label{eq:mixing-Liv}
\bigg|\int a\circ\Phi^t\, b\,\, \ud\mu -\int a\,\, \ud\mu\int b\,\, \ud\mu\bigg|\leq C |a|_{\alpha}|b|_{\alpha}\ue^{-\gamma t}\,\, .
\end{equation}
Quantitative results on the decay of correlations for Anosov 
flows are rather recent, the main results prior to 
\cite{Liv03} were obtained by Chernov \cite{Che98} and 
Dolgopyat \cite{Dol98}, see the introduction of \cite{Liv03} for 
more details on the history of this problem. Since the restriction 
of a Hamiltonian flow to an energy shell is a contact flow, 
the result of Liverani applies to the systems we are interested in.

We want to extend the result of Liverani to the case that one of the 
functions in the correlation integral is a density concentrated 
on a smooth submanifold. Such results have been obtained previously for 
goedesic flows on manifolds of negative curvature with certain measures 
concentrated on the unstable manifolds by Sinai and Chernov. 
Sinai showed in \cite{Sin95} that mixing holds and Chernov, \cite{Che97}, 
showed that the correlations decay at least like 
$\ue^{-\gamma \sqrt{t}}$. On manifolds of constant negative 
curvature  Eskin and McMullen, \cite{EskMcm93}, derived mixing if one of the 
functions is concentrated on certain submanifolds. They reduced this 
to the classical mixing results for functions by using the hyperbolicity 
of the flow. We will follow their approach, where the only additional 
difficulty coming in is that the stable foliation is no longer smooth 
but only  H{\"o}lder continuous if the curvature is no longer constant. 
To overcome this we use the absolute continuity property of the stable 
foliation.   
 
In the following we will assume that  non-vanishing
smooth densities $\sigma_{\Lambda}$ and $\sigma_{\Gamma}$ have  been 
fixed on the submanifolds 
$\Lambda$ and $\Gamma$, so that every density can be written 
as $\sigma=\hat{\sigma}\sigma_{\lambda}$ or $\sigma=\hat{\sigma}\sigma_{\Gamma}$. We say then that 
$\sigma\in C^{\alpha}(\Lambda)$ if $\hat{\sigma}\in C^{\alpha}(\Lambda)$ 
and analogously 
$\sigma\in C^{\alpha}(\Gamma)$ if $\hat{\sigma}\in C^{\alpha}(\Gamma)$.


\begin{theorem}\label{thm:mixing}
Let $S$ be a symplectic manifold of dimension $2d$, 
and $\Phi^t: S\to S$ be a Hamiltonian flow
on $S$ with Hamilton-function $H\in C^{\infty}(S)$. Denote by 
$\Sigma_E:=\{z\in S\, ; H(z)=E\}$ the energy shell with energy $E$ and 
by $\ud \mu_E$ the Liouville measure on $\Sigma_E$.  
Assume $\Sigma_E$ is compact and connected, and $\Phi^t$ is 
Anosov on $\Sigma_E$ and the stable foliation  has H{\"o}lder exponent 
$\alpha$.

\begin{itemize}
\item[(i)] Let $\Lambda\subset \Sigma_E$ be a d-dimensional submanifold 
which is transversal to the stable foliation of $\Sigma_E$ except 
on a subset of codimension at least $1$. Then there exist 
$\gamma_1>0$ and for every  
density $\sigma\in C^{\alpha}_0(\Lambda)$  a constant $C_1$ such that 
for 
every function $a\in C^{\alpha}(\Sigma_E)$ we have 
\begin{equation}
\bigg| \int_{\Lambda} a\circ \Phi^t \, \sigma -\int_{\Sigma_E} a\, \ud \mu_E 
\int_{\Lambda} \sigma \bigg|\leq C_1 |a|_{\alpha} 
\ue^{-\gamma_1 t} 
\end{equation}
\item[(ii)] Let $\Gamma\subset \Sigma_E$ be a $(d-1)$-dimensional submanifold 
which is transversal to the weakly-stable foliation of $\Sigma_E$, except 
on a subset of codimension at least $1$. Then there exist 
$\gamma_2>0$ and for every  
density $\sigma\in C^{\alpha}_0(\Gamma)$ a $C_2$ such that for 
every function $a\in C^{\alpha}(\Sigma_E)$ we have 
\begin{equation}
\bigg| \int_{\Gamma} a\circ \Phi^t \, \sigma -\int_{\Sigma_E} a\, \ud \mu_E 
\int_{\Gamma} \sigma \bigg|\leq C_2 |a|_{\alpha} 
\ue^{-\gamma_2 t} 
\end{equation}
\item[(iii)] Let $\Lambda\subset S$ be a d-dimensional submanifold and assume that 
the flow is Anosov on all $\Sigma_E$ with $\Sigma_E\cap\Lambda\neq \emptyset$.  
Assume furthermore that $\Lambda\cap\Sigma_E$ is transversal to the weakly stable 
foliation of $\Sigma_E$ for all 
$E$, except on a subset of codimension at least one.
Then there exist 
$\gamma_3>0$ and for  for every  
density $\sigma\in C^{\alpha}_0(\Lambda)$ a constant $C_3$ such that for 
every function $a\in C^{\alpha}_0(S)$ we have 
\begin{equation}
\bigg| \int_{\Lambda} a\circ \Phi^t \, \sigma -\int\int_{\Sigma_E} a\, \ud \mu_{E} 
\int_{\Lambda\cap\Sigma_E} \sigma_E\, \, \ud E \bigg|\leq C_3 |a|_{\alpha} 
\ue^{-\gamma_3 t}\,\, , 
\end{equation}
where $\sigma_E$ is a density on $\Lambda\cap\Sigma_E$ defined by 
$\sigma =\sigma_E \otimes \abs{\ud E}$
\end{itemize} 
\end{theorem}


\begin{proof}
In order to prove $(i)$, we will relate the behaviour of 
\begin{equation}\label{eq:sing-cor-proof}
\int_{\Lambda}a\circ\Phi^t\, \sigma
\end{equation}
to the behaviour of the standard correlation function 
\begin{equation}\label{eq:reg-cor-proof}
\int_{\Sigma_E} a\circ\Phi^t \rho \, \ud \mu_E 
\end{equation}
where $\rho\in C^{\alpha}(\Sigma_E)$ is supported in a neighbourhood of $\Lambda$. 
The heuristic idea is that since a neighbourhood of $\Lambda$ converges 
exponentially fast along the stable manifolds to $\Lambda$, the integral 
\eqref{eq:reg-cor-proof} will become close to the integral 
\eqref{eq:sing-cor-proof} for appropriately chosen $\rho$. 
But to \eqref{eq:reg-cor-proof} we can then apply the result \eqref{eq:mixing-Liv}  
by Liverani.

We will formalise this idea now and treat first the case that $\Lambda$ 
is transversal to the stable foliation. 
By using a partition of unity we can assume that the support of 
$\sigma$ is in a small compact set $\Lambda_0 \subset \Lambda$, such that there 
is a neighbourhood $\hat{\Lambda}_0\subset \Sigma_E$ of $\Lambda_0$ in $\Sigma_E$ 
in which we can choose coordinates 
$(x,y)\in U\times W\subset\R^d\times\R^{d-1}$ with the property that 
$\Lambda= \{(x,0),x\in U\}$ and $W^{s}(x)=\{(x,y);y\in W\}$. 
This is where we use the transversality assumption. Notice that since the 
stable foliation is usually only H{\"o}lder continuous, 
the transformation to this  coordinate system is 
only H{\"o}lder continuous, too. Now the 
absolute continuity of 
the stable foliation means that there is a measurable function 
$\delta_x(y)$ which depends measurably on $x$ and satisfies 
$1/C<\delta_x(y)<C$ for some $C>0$ and all $(x,y)\in U\times W$, such that 
\begin{equation}
 \int_{\Sigma_E} a\circ\Phi^t \rho \, \ud \mu_E =
\int_{U}\int_W \rho(x,y) a\circ\Phi^t(x,y) \delta_x(y)\, \ud y\ud x\,\, ,
\end{equation}
where we have assumed that $\rho$ is supported in $U\times W$, see 
\cite[Chapter 6.2]{BriStu02}. We will now assume that 
$\rho$ can be chosen to be in $C^{\alpha}(\Sigma_E)$ and such that 
\begin{equation}\label{eq:rho-normalization}
\int_W \rho(x,y)\delta_x(y)\, \ud y=\hat{\sigma}(x)
\end{equation}
where $\sigma(x)=\hat{\sigma}(x)\ud x$, we will show 
below that this is possible. By H{\"o}lder continuity we get  now 
\begin{equation}
|a\circ\Phi^t(x,y)-a\circ\Phi^t(x,0)|\leq C|a|_{\alpha} d(\Phi^t(x,y),\Phi^t(x,0))^{\alpha}\leq C'|a|_{\alpha} \ue^{-\alpha\gamma t}\, \, ,
\end{equation}
since the flow is contracting along the stable leaves, i.e.,  
$d(\Phi^t(x,y),\Phi^t(x,0))\leq C \ue^{-\gamma t}$ for some constants 
$C,\gamma>0$. 
Therefore we obtain with \eqref{eq:rho-normalization}
\begin{equation}
\begin{split}
\bigg|\int_{U}\int_W &\rho(x,y) a\circ\Phi^t(x,y) \delta_x(y)\, \ud y\ud x
-\int_{U}\int_W \rho(x,y) a\circ\Phi^t(x,0) \delta_x(y)\, \ud y\ud x\bigg|\\
&\leq C'|a|_{\alpha} \int_U |\hat{\sigma}(x)|\, \ud x\, \ue^{-\alpha\gamma t}
\end{split}
\end{equation}
and 
\begin{equation}
\int_{U}\int_W \rho(x,y) a\circ\Phi^t(x,0) \delta_x(y)\, \ud y\ud x
=\int_U a \circ\Phi^t(x,0)\hat{\sigma}(x)\, \ud x= \int_\Lambda a \circ\Phi^t\, \sigma\,\, .
\end{equation}
On the other hand we have by \eqref{eq:mixing-Liv}
\begin{equation}
\bigg|\int_{\Sigma_E} a\circ\Phi^t \rho \, \ud \mu_E
-\int_{\Sigma_E} \rho \, \ud \mu_E\int_{\Sigma_E} a\, \ud \mu_E\bigg|
\leq C |a|_{\alpha}|\rho|_{\alpha} \ue^{-\gamma' t}
\end{equation}
and by \eqref{eq:rho-normalization} 
\begin{equation}\label{eq:rho-sigma-estimates}
\int_{\Sigma_E} \rho \, \ud \mu_E
=\int_{\Lambda}\sigma \,\, , \quad  |\rho|_{\alpha}\leq C_{\Lambda} \,
|\hat{\sigma}|_{\alpha}
\end{equation} 
so finally we get 
\begin{equation}
\bigg|\int_{\Lambda} a\circ\Phi^t \sigma -\int_{\Lambda} \sigma \, 
\int_{\Sigma_E} a\,\, \ud \mu_{E}\bigg| \leq C (|\sigma|_{\alpha}+||\sigma||_{L^1(\Lambda)})|a|_{\alpha} \ue^{-\gamma t}
\end{equation}

We still have to check that one can choose a $\rho\in C^{\alpha}$ which 
satisfies \eqref{eq:rho-normalization}. Set $\rho(x,y)=\rho_1(x)\rho_2(x,y)
\hat{\sigma}(x)$ with $\rho_2(x,y)>0 $ on $\Lambda_0$, H{\"o}lder  and supported in $\hat{\Lambda}_0$, 
and set 
\begin{equation}
\rho_1(x)=\bigg(\int_{W}\rho_2(x,y)\delta_x(y)\,\ud y\bigg)^{-1}
\end{equation}
on $\Omega$. Then $\rho_1$ is H{\"o}lder, since the foliation $W^s(x)$ is 
H{\"o}lder, and therefore $\rho$ is H{\"o}lder too.   This  completes 
the proof of $(i)$ in case the manifolds are transversal. 

We will now extend this result to the non-transversal case. 
Let $\Lambda_{sing}=\{x\in\Lambda ; \dim T_x\Lambda\cap T_xW^s(x)\geq 1\}$ 
be the set of point on $\Lambda$ where the intersection is not transversal, 
and define $\Lambda_{sing,\varepsilon}:=
\{x \in \Lambda ; d(x,\Lambda_{sing})\leq \varepsilon \}$. 
Choose $\varphi_{\varepsilon}\in C^{\alpha}(\Lambda)$ with 
$\supp\varphi_{\varepsilon}\subset \Lambda_{sing,\varepsilon}$ and 
$\varphi_{\varepsilon}\equiv 1$ on $\Lambda_{sing,\varepsilon/2}$. 
Then 
\begin{equation}
\biggabs{\int_{\Lambda}\varphi_{\varepsilon}a\circ \Phi^t 
\abs{\sigma(\psi)}^2 }\leq C\abs{a}\varepsilon^{d-d_{sing}} \,
\, . 
\end{equation}
where $d_{sing}$ is the dimension of $\Lambda_{sing}$. 

To the integral $\int_{\Lambda}(1-\varphi_{\varepsilon})
a\circ \Phi^t \abs{\sigma(\psi)}^2$ we can apply the previous results, 
we only have to pay attention to the $\varepsilon$-dependence 
of the constants. The second estimate in 
\eqref{eq:rho-sigma-estimates} has to be refined. By the definition 
of $\rho$ we have $\abs{\rho(1-\varphi_{\varepsilon})}_{\alpha}\leq \abs{\rho_1(1-\varphi_{\varepsilon})}_{\alpha}\abs{\rho_2}_{\alpha}\abs{\hat{\sigma}}_{\alpha}$ and since the Jacobian $\delta_y(x)$ becomes degenerate 
when $x$ approaches $\Lambda_{sing}$  we get 
\begin{equation}
\abs{\rho_1(1-\varphi_{\varepsilon})}_{\alpha} 
\leq C \varepsilon^{-\gamma'}
\end{equation}
where $\gamma' >0$ depends on $\alpha$ and $d_{sing}$. 
Collecting the estimates yields  
\begin{equation}\label{eq:rho-sigma-estimates-gen}
\bigg|\int_{\Lambda} a\circ\Phi^t \sigma -\int_{\Lambda} \sigma \, 
\int_{\Sigma_E} a\,\, \ud \mu_{E}\bigg| \leq C \varepsilon^{-\gamma'}(|\sigma|_{\alpha}+||\sigma||_{L^1(\Lambda)})|a|_{\alpha} \ue^{-\gamma t}
+C'\abs{a}\varepsilon^{d-d_{sing}}
\end{equation}
and choosing $\varepsilon=\ue^{-\gamma'' t}$ with $\gamma''=\gamma/(\gamma'+(d-d_{sing}))$ gives 
\begin{equation}
\bigg|\int_{\Lambda} a\circ\Phi^t \sigma -\int_{\Lambda} \sigma \, 
\int_{\Sigma_E} a\,\, \ud \mu_{E}\bigg| \leq C (|a|_{\alpha}+\abs{a}) \
\ue^{-\gamma_1 t}
\end{equation}
with $\gamma_1=\gamma(d-d_{sing})/(\gamma'+(d-d_{sing}))$.

The proof of $(ii)$ is based on $(i)$.  Define for some $\delta>0$ 
$\Lambda :=\bigcup_{|t|< \delta} \Phi^t(\Gamma)\subset\Sigma_E$, then 
$\Lambda$ is transversal to the stable foliation except on a subset 
of codimension at least one. If 
$s\in U\subset \R^{d-1}$ are local coordinates on 
$\Gamma$, then $(r,s)$ $|r|< \delta$ are local coordinates on 
$\Lambda$. Let $\rho$ be a smooth function with compact support 
in $|r|<\delta$, $\int \rho(r)\,\, \ud r =1$, and define $\rho_{\varepsilon}(r):=\frac{1}{\varepsilon}
\rho(\varepsilon r) $. If we write $\sigma =\hat{\sigma}(s)\, \ud s$ and 
$\sigma_{\varepsilon}:=\hat{\sigma}(s)\rho_{\varepsilon}(r)\ud s\ud r$, we have 
\begin{equation}
\begin{split}
\bigg| \int_{\Gamma}  a\circ\Phi^t\,\sigma -\int_{\Lambda}a\circ\Phi^t\, \sigma_{\varepsilon}
\bigg|
&= \bigg| \int_{U}a(t,s)\,\hat{\sigma}\,\,\ud s -\int_{U}\int_{\R}a(r+t,s)\, \rho_{\varepsilon}(r)\hat{\sigma}(s)\,\, \ud r\ud s\bigg|\\
&\leq  \int_{U}\int_{\R}\rho_{\varepsilon}(r)|a(t,s)-a(r+t,s)|\,\, \ud r\,\hat{\sigma}(s)\,\,\ud s  
\end{split}
\end{equation}
but 
\begin{equation}
\int_{\R}\rho_{\varepsilon}(r)|a(t,s)-a(r+t,s)|\,\, \ud r
=\int_{\R}\rho(r)|a(t,s)-a(\varepsilon r+t,s)|\,\, \ud r
\leq C |a|_{\alpha} \varepsilon^{\alpha}
\end{equation}
and therefore 
\begin{equation}
\bigg| \int_{\Gamma} a\circ\Phi^t\,\sigma -\int_{\Lambda}a\circ\Phi^t\, \sigma_{\varepsilon}
\bigg|\leq C ||\sigma||_{L^1(\Sigma)}|a|_{\alpha} \varepsilon^{\alpha}\,\, .
\end{equation}
On the other hand with $|\sigma_{\varepsilon}|_{\alpha}\leq C |\sigma |_{\alpha} \varepsilon^{\alpha-1}$ and $||\sigma_{\varepsilon}||_{L^1(\Lambda)}=||\sigma||_{L^1(\Gamma)}$
we obtain from $(i)$ that 
\begin{equation}
\bigg|\int_{\Lambda}a\circ\Phi^t\, \sigma_{\varepsilon}
-\int_{\Sigma_E} a\,\, \ud \mu_E \, \int_{\Gamma}\sigma\bigg|\leq 
C |a|_{\alpha}(|\sigma|_{\alpha}\varepsilon^{\alpha -1} +||\sigma||_{L^1(\Gamma)})
\ue^{-\gamma_1 t}\,\, .
\end{equation}
If we now choose $\varepsilon=\ue^{-\gamma' t}$ with $\gamma'>0$ and 
$(1-\alpha)\gamma'>\gamma_1$, the proof of $(ii)$ is complete.

Part $(iii)$ then follows immediately  by writing 
\begin{equation}
\int_{\Lambda}a\circ\Phi^t \sigma =\int \int_{\Lambda\cap\Sigma_E} a\circ\Phi^t \sigma_E\, \ud E
\end{equation}
and applying $(ii)$ to the  integral over $\Lambda\cap\Sigma_E$ on the right hand side. 
 
\end{proof}

Theorem \ref{thm:main-anosov} is now a 
straightforward consequence of 
Proposition \ref{prop:reduction} and Theorem \ref{thm:mixing}.

Let us end this section by discussing the meaning of the transversality condition. 
Let us first look at the example that $\Lambda$ is the stable manifold of an 
periodic orbit $\gamma$ with period $T_{\gamma}$. 
Let $(r,x)\in S^1 \times \R^{d-1}$ be coordinates on $\Lambda$ 
such that $\gamma$ is given by $x=0$ and $\Phi^t(r,0)=(r+t \mod T_{\gamma} ,0)$, then 
\begin{equation}
\int_{\Lambda}a \circ \Phi^t\, \sigma = 
\int_0^{T_{\gamma}}\int_{\R^{d-1}} a(r+t,x(t)) \hat{\sigma}(r,x) \,\, \ud r \ud x\,\, .
\end{equation}
With $\abs{a(r+t,x(t))-a(r+t,0)}\leq C \ue^{-\gamma t}$ and by inserting 
the Fourier series $a(r,0)=\sum_{k\in \Z}a_k \ue^{\frac{2\pi}{T_{\gamma}} \ui k r}$ 
we obtain 
\begin{equation}
\int_{\Lambda}a \circ \Phi^t\, \sigma = 
\sum_{k\in \Z}a_k \tilde{\sigma}_k \ue^{\frac{2\pi}{T_{\gamma}} \ui r}
+O(\ue^{-\gamma t})
\end{equation}
with 
$\tilde{\sigma}_k=\int_0^{T_{\gamma}}\int_{\R^{d-1}}\hat{\sigma}(r,x)\, \ud x\, 
\ue^{\frac{2\pi}{T_{\gamma}} \ui k r}\,\, \ud r$. 
So in this case we do not get convergence for large times, 
and together with Proposition \ref{prop:reduction} this gives \eqref{eq:per-orbit}.  This example shows that some condition on the position of 
$\Lambda$ with respect to the 
stable foliation is necessary.

\section{Integrable systems}\label{sec:integrable}

In this section we give the proof of Theorem \ref{thm:limit-int-syst} 
and discuss the situation for integrable systems. 

\begin{proof}[Proof of Theorem \ref{thm:limit-int-syst}] 
By Proposition \ref{prop:reduction} we have to study the behaviour of 
\begin{equation}
\int_{\Lambda} \sigma(a)\circ\Phi^t \, 
\abs{\sigma(\psi)}^2\,\, ,
\end{equation}
for large $t$. Assume first that $\Lambda$ is a part of 
an invariant torus.  In action angle coordinates $(I,x)\in U\times \bT^d$ 
it is the given by $\Lambda=\{(I,x), x\in V\subset \bT^d\}$, 
so we get 
\begin{equation}
\int_{\Lambda} \sigma(a)\circ\Phi^t \, 
\abs{\sigma(\psi)}^2
=\int_{\bT^d} \sigma(a)(I,x+t\omega(I)) \abs{\rho(x)}^2\,\, \ud x\,\, .
\end{equation}
If we insert now for $\sigma(a)(I,x)$ its Fourier series in $x$ we obtain
\begin{equation}
\int_{\Lambda} \sigma(a)\circ\Phi^t \, 
\abs{\sigma(\psi)}^2
=\sum_{m\in\Z^d} \sigma(a)_m(I)
\int_{\bT^d} \ue^{\ui\la x,m\ra}
\abs{\rho(x)}^2\,\, \ud x \, \ue^{\ui t\la \omega(I),m\ra}\,\, ,
\end{equation}
which is equation \eqref{eq:int-lambda-torus} in Theorem \ref{thm:limit-int-syst}.

In order to prove equation \eqref{eq:int-lambda-transversal} we notice that  the 
 transversality assumption on $\Lambda$ with respect to the foliation in 
invariant tori implies that in action angle coordinates 
$(I,x)\subset U\times V$ $\Lambda$ can locally represented by a 
generating function 
\begin{equation}
\Lambda=\{(I,\varphi'(I))\,, \,   I\in U\}.
\end{equation}
Therefore we have 
\begin{equation}
\int_{\Lambda} \sigma(a)\circ\Phi^t \, 
\abs{\sigma(\psi)}^2=\int_{U}\sigma(a)(I,\varphi'(I)+t\omega(I))\abs{\hat{\rho}(I)}\,\, \ud I\,\, ,
\end{equation}
and inserting for $\sigma(a)$ again the Fourier expansion in $x$ leads to
\begin{equation}
\int_{\Lambda} \sigma(a)\circ\Phi^t \, 
\abs{\sigma(\psi)}^2=\sum_{m\in\Z^d}\int_{U}\sigma(a)_m(I)
\ue^{\ui \la m,\varphi'(I)\ra}\ue^{\ui t\la m,\omega(I)\ra}\abs{\hat{\rho}(I)}\,\, \ud I\,\, ,
\end{equation}

The non-degeneracy condition 
$\omega '(I)\neq 0$  implies that there exist a constant $C>0$ 
\begin{equation}
\abs{\nabla_I \la \omega(I), m\ra} 
\geq C \abs{m}\,\, ,
\end{equation}
for all $I\in \supp \hat{\rho}$. 
Now by the non-stationary phase estimates, see, e.g., 
\cite[Theorem 7.7.1]{Hor90}, on gets 
\begin{equation}
\biggabs{\int_{U} \sigma(a)_m(I)\ue^{\ui \la m,\varphi'(I)\ra}\ue^{\ui t\la m,\omega(I)\ra}\abs{\hat{\rho}(I)}\,\, \ud I}\leq 
C \abs{m}\abs{\sigma(a)_m}_1\abs{\rho}_1 \, \frac{1}{1+\abs{t}}
\end{equation}
for $m\neq 0$. And therefore we finally obtain 
\begin{equation}
\int_{\Lambda} \sigma(a)\circ\Phi^t \, 
\abs{\sigma(\psi)}^2
=\int_{U} \sigma(a)_0(I)\abs{\hat{\rho}(I)}^2 \,\, \ud x 
+O(1/t)
\end{equation}
and so the proof of Theorem  \ref{thm:limit-int-syst} is complete.
\end{proof}

There are a couple of directions in which one probably can extend and improve 
Theorem \ref{thm:limit-int-syst}.  We have only studied the two extreme cases 
of the position of $\Lambda$ relative to the foliation into invariant tori. Certainly 
the transversal case is (locally) generic, but the case that the intersections are clean 
can be studied without much additional effort, one would expect an oscillatory behaviour in 
this case. It appears as well to be very interesting to investigate the behaviour of the 
time evolution close to singularities of the foliation into invariant tori.  

Another direction where one can generalise some of the results is to more general classes 
of systems. Namely by using normal forms around invariant tori in general 
system on can extent 
the result $(i)$ to that case. Such invariant tori occur typically in 
situation described by 
KAM theory, e.g., for perturbed integrable systems, and close to elliptic orbits.

\vspace{0.5cm}
\noindent {\bf{Acknowledgements}}.
This work is a result of my research during my stays at the SPhT in Saclay/Paris and the 
MSRI in Berkeley, I would like to thank St{\'e}phane Nonnemacher, 
Boris Gutkin and Andre Voros for interest and support. 
This work has been fully supported by the European Commission under the Research Training Network (Mathematical Aspects of Quantum Chaos) n° HPRN-CT-2000-00103 of the IHP Programme.



\def\cprime{$'$}
\providecommand{\bysame}{\leavevmode\hbox to3em{\hrulefill}\thinspace}
\providecommand{\MRhref}[2]{%
  \href{http://www.ams.org/mathscinet-getitem?mr=#1}{#2}
}
\providecommand{\href}[2]{#2}

\end{document}